\documentclass[12pt,notitlepage]{article}

\usepackage{amsmath,amssymb}
\usepackage{graphicx}

\usepackage{color}

\def\ignore#1{{}}

\let\oldtheequation=\theequation
\def\doteqs#1{\setcounter{equation}{0}            
	\def\theequation{{#1}.\oldtheequation}}
\newcounter{sxn}
\def\sx#1{\addtocounter{sxn}{1} \vskip 1.cm  \goodbreak
	\noindent{\large\bf\leftline{\thesxn.~~#1}} \nobreak \vskip -.5cm}
\def\sxn#1{\sx{#1} \doteqs{\thesxn}}

\newcounter{axn}

\newdimen\mybaselineskip
\newcommand{\beeq}{\begin{equation}}
	\newcommand{\eneq}{\end{equation}}
\newcommand{\beqn}{\begin{eqnarray}}
	\newcommand{\eeqn}{\end{eqnarray}}

\newcommand{\ba}{\begin{array}}
	\newcommand{\ea}{\end{array}}

\newcommand{\be}{\begin{equation}}
	\newcommand{\ee}{\end{equation}}
\newcommand{\bea}{\begin{eqnarray}}
	\newcommand{\eea}{\end{eqnarray}}
\newcommand{\beal}{\setcounter{letter}{1} \begin{eqnarray}}
	\newcommand{\eeal}{\addtocounter{equation}{1} \end{eqnarray}}

\newcommand{\larrow}{\,\,\,\,\hbox to 30pt{\rightarrowfill}
	\,\,\,\,}
\newcommand{\slarrow}{\,\,\,\hbox to 20pt{\rightarrowfill}
	\,\,\,}

\def\la{\raise.16ex\hbox{$\langle$}\lower.16ex\hbox{}  }
\def\ra{\, \raise.16ex\hbox{$\rangle$}\lower.16ex\hbox{} }

\def\psibar{ \psi \kern-.65em\raise.6em\hbox{$-$} \lower.6em\hbox{} }
\def\psibarb{ \psi \kern-.65em\raise.6em\hbox{$-$}  }

\begin{document}

\thispagestyle{empty}

\baselineskip=12pt



\vspace*{3.cm}

\begin{center}  
{\LARGE \bf  Calculating quasinormal modes of extremal 
	 and non-extremal Reissner-Nordstr\"om black 
	 holes with the continued fraction method}
\end{center}

\baselineskip=14pt

\vspace{1cm}
\begin{center}
{\bf  Ramin G. Daghigh$^1$, Michael D. Green$^2$, and  Jodin C. Morey$^3$}
\end{center}

\vspace{0.25 cm}
\centerline{\small \it $^1$ Natural Sciences Department, Metropolitan State University, Saint Paul, Minnesota, USA 55106}
\vskip 0 cm
\centerline{} 

\centerline{\small \it $^2$ Mathematics and Statistics Department, Metropolitan State University, Saint Paul, Minnesota, USA 55106}
\vskip 0 cm
\centerline{} 

\centerline{\small \it $^3$ School of Mathematics, University of Minnesota, Minneapolis, Minnesota, USA 55455}
\vskip 0 cm
\centerline{}

\vspace{1cm}
\begin{abstract}
We use the numerical continued fraction method to investigate quasinormal mode spectra of extremal and non-extremal Reissner-Nordstr\"om black holes in the low and intermediate damping regions.  
In the extremal case, we develop  techniques that significantly expand the calculated spectrum from what had previously appeared  in the literature.
This allows us to determine the asymptotic behavior of the extremal spectrum in the high damping limit, where there are conflicting published results.
Our investigation further supports the idea that the extremal limit of the non-extremal case, where the charge approaches the mass of the black hole in natural units, leads to the same vibrational spectrum  as in the extremal case despite the qualitative differences in their topology. In addition, we numerically explore the quasinormal mode spectrum for a Reissner-Nordstr\"om black hole in the small charge limit. 

\baselineskip=20pt plus 1pt minus 1pt
\end{abstract}

\newpage

\sxn{Introduction}

\vskip 0.3cm

Natural vibrational modes, also called quasinormal modes (QNMs), of black holes have attracted a great deal of attention since the first detection of gravitational waves emitted from a binary black hole merger\cite{GW-detection}.   There also have been efforts in establishing a link between QNMs and the quantum structure of a black hole spacetime.  See for example \cite{Hod, Maggiore, Babb}.


In this paper we focus on the QNMs of extremal and non-extremal Reissner-Nordstr\"om (RN)  black holes with charge $q$ and mass $M$.  Extremal black holes, when $q=M$ using units where $G=c=k_e=1$, hold an important and controversial status in black hole physics.  Traditionally these black holes were believed to be a limiting case of non-extremal black holes\cite{Misner}.  This traditional view was challenged by Hawking and Horowitz in \cite{Hawking-Horowitz} based on the fact that the topology of the extremal and non-extremal black holes have qualitative differences.  Due to these differences, the authors of \cite{Hawking-Horowitz} and \cite{Teitelboim} argued that extremal RN black holes have zero entropy with no definite temperature despite having a non-zero horizon area.  Extremal black holes are also important in the context of supergravity theories\cite{Supergravity}.   

In \cite{Onozawa1}, Onozawa {\it et al.}\ numerically computed the least-damped modes  for four dimensional extremal RN black holes.  They found that the QNM spectra of gravitational waves with multipole index $\ell$ and electromagnetic waves with multipole index $\ell-1$ coincide\footnote{Electromagnetic and gravitational waves in the RN black hole are coupled.  Here, electromagnetic (gravitational) refers to perturbations that reduce to pure electromagnetic (gravitational) in the zero charge limit.}.  Based on this observation, the authors of \cite{Onozawa2}  conjectured that the modes of different perturbations can be matched because of the supersymmetry transformations in the extremal solution.  
In \cite{Berti1}, Berti was able to extend the numerical calculation of \cite{Onozawa1} to include higher damped modes.  Berti's results show that the extremal RN QNM spectrum has a similar pattern to the Schwarzschild case.  As the damping increases the oscillation frequency seems to approach the same constant value as in Schwarzschild black holes, i.e.
\beeq
\omega_R \mathop{\longrightarrow}_{|\omega_I|\to\infty}\frac{\ln(3)}{8\pi  M} ~,
\label{ln3-extreme}
\eneq
where $\omega_R$ represents the real part of the vibration frequency of the QNM, $\omega$, and $\omega_I$ is the imaginary part that determines the damping rate.
This result seems to be compatible with the extremal limit, where $q\rightarrow M$,  of the non-extremal RN QNM spectrum which was derived in \cite{Andersson} for the highly damped limit ($|\omega_I|\to\infty$). Unfortunately the numerical method in \cite{Berti1} is unstable after approximately twenty roots.  Therefore, one cannot
verify the limit (\ref{ln3-extreme}).  It is suggested in \cite{Andersson} that since the topology of the Stokes/anti-Stokes lines for the extremal RN black hole is different than the non-extremal case, the QNMs for these black holes would require a separate analysis.  Following this suggestion, the authors of \cite{Das1} and \cite{Natario-S} have attempted to explicitly calculate the highly damped QNM frequency of extremal RN black holes using the monodromy method of \cite{Motl2}. Neither of the analytic results in \cite{Das1} and \cite{Natario-S} match the QNM spectrum of the extremal limit of the non-extremal case.  Therefore, they contradict the numerical results provided by Berti in \cite{Berti1}.    In \cite{DG-extremal}, two of us used the topology of anti-Stokes lines presented in \cite{Natario-S} to show that in the higly damped limit a particular path along the anti-Stokes lines will lead to results in agreement with Berti's numerical calculation in \cite{Berti1} and with the extremal limit of the non-extremal case derived in \cite{Andersson}.  A  geometric treatment of the extremal and near extremal cases in \cite{Rodrigo} also leads to similar conclusions.    
In this paper, we extend the numerical results of Berti to higher overtone QNMs to show that the QNM spectrum of the extremal case matches the extremal limit.


Another issue that we examine in this paper is about the transition between intermediate and high damping regions of the QNM spectrum of RN black holes in the small charge limit ($q \rightarrow 0$).
It was shown first by Motl and Neitzke\cite{Motl2} and then confirmed by Andersson and Howls\cite{Andersson} that for RN black holes, the real part of the highly damped QNM frequency in four dimensions approaches $\ln(5)/(8\pi M)$ as the charge goes to zero. The general validity of this result was verified by Natario and Schiappa\cite{Natario-S}, who calculated RN highly damped spectra in arbitrary spacetime dimensions. The apparent contradiction between the zero charge limit of the RN case, and the Schwarzschild value was explained heuristically by the authors of \cite{Andersson}. They noted, that, while  $\ln(5)/(8\pi M)$ represents the correct RN result for very high damping, one expects an intermediate range of damping in which one finds the Schwarzschild value of $\ln(3)/(8\pi M)$ for $\omega_R$. Order of magnitude arguments suggest this range in four spacetime dimensions to be:
\beeq
1<<|\omega|^2M^2 << M^8/q^8
\label{Eq: range-small-q}
\eneq
In \cite{Daghigh-RN}, the authors use a combination of analytical and numerical techniques to analyze the limit of large but finite damping where the Schwarzschild limit is approached. For four and five spacetime dimension, the authors of  \cite{Daghigh-RN} explicitly calculate the spectrum in the transition region from $\ln(3+4\cos({d-3 \over 2d-5}\pi))/(8\pi M)$, in $d$ spacetime dimensions, for very large damping to the Schwarzschild value of $\ln(3) /(8\pi M)$. Based on this work, the real frequency does not interpolate smoothly between the two values. Instead there is a critical value of the damping at which the Stokes/anti-Stokes lines change topology and the real part of the frequency dips to zero. This behavior seems to resemble the algebraically special frequencies that mark the onset of the high damping regime.  


The paper is organized as follows.  In Section 2, we describe the general formalism and the continued fraction method for the non-extremal case.  In Section 3, we describe numerical methods to address the extremal case.  In section 4, we report and discuss the results.  Finally, in Section 5 we end the paper with some concluding remarks.

\sxn{General Formalism}
\vskip 0.3cm

Various classes of non-rotating black hole metric perturbations  are governed generically by a Schr$\ddot{\mbox o}$dinger-like wave equation of the form
\beeq
{d^2\psi \over dr_*^2}+\left[ \omega^2-V(r) \right]\psi =0 ~,
\label{Schrodinger}
\eneq
where $V(r)$ is the QNM potential.  In this paper, we assume the perturbations depend on time as $e^{-i\omega t}$.  Consequently, in order to have damping, the imaginary part of $\omega$ must be negative.  
The Tortoise coordinate $r_*$ is defined by 
\beeq
dr_* ={dr \over f(r) }~,
\label{tortoise}
\eneq
where the metric function, $f(r)$, for the Reissner-Nordstr\"om spacetime is given by 
\beeq
f(r)=1-{2M \over r}+ \frac{q^2}{r^2}~.
\label{function f}
\eneq
Combining the above two equations gives us the tortoise coordinate
\beeq
r_*=r+ \frac{r_+^2}{r_+-r_-}\ln(r-r_+)-\frac{r_-^2}{r_+-r_-}\ln(r-r_-)~
\label{Eq:Tortoise}
\eneq
where $r_\pm =(M\pm \sqrt{M^2-q^2})$ are the roots of the metric function that determine the locations of the inner (Cauchy) and outer (event) horizons.  
$M$ is the ADM mass of the black hole and $q$ is the charge. 
The effective potential\footnote{Here we only consider axial perturbations since the polar perturbations can be obtained from axial perturbations using the Chandrasekhar transformation \cite{Chandra}.} in Eq.\ (\ref{Schrodinger}) is 
\beeq
V_s(r) = f(r)\left[ \frac{\ell (\ell + 1)}{r^2} - \frac{\beta_s}{r^3} + \frac{4 q^2}{r^4} \right],
\label{Vr}
\eneq
where $\ell$ is the multipole number, $s=1$ is the spin for perturbations that reduce to pure electromagnetic as $q\rightarrow0$ (hereafter ``electromagnetic'' for brevity) and $s=2$ is the spin for perturbations that reduce to pure axial gravitational perturbations as $q\rightarrow0$ (hereafter ``gravitational'' for brevity), and
\beeq
 \beta_s= 3M+(-1)^s \sqrt{9M^2 +4q^2(\ell-1)(\ell+2)}.
\label{Vr}
\eneq

The effective potential is zero at the event horizon ($r \rightarrow r_+$) and at spatial infinity ($r \rightarrow \infty$).  
The QNMs are obtained using the boundary conditions in which the asymptotic behavior of the solutions is
\beeq
\psi(r) \approx \left\{ \begin{array}{ll}
	e^{-i\omega r_*}  & \mbox{as $r_* \rightarrow -\infty$ }\\
	e^{i\omega r_*}   & \mbox{as $r_* \rightarrow \infty$ }
\end{array}
\right.        
\label{asymptotic}
\eneq 
representing in-going waves at the event horizon and outgoing waves at infinity.

We can write the solution to the wave equation (\ref{Schrodinger}), with the desired behavior at the boundaries, in the form
\beeq
\psi(r) =  \frac{r_+ e^{-2 i \omega r_+} (r_+-r_-)^{-2 i \omega -1} (r-r_-)^{1+i \omega} e^{i \omega r}}{r} 
  u^{-i \omega r_+^2/(r_+-r_-)}   \sum_{n=0}^{\infty}a_n  u^{n
} ~,
\label{eq-series}
\eneq
where $u=(r-r_+)/(r-r_-)$.
Inserting (\ref{eq-series}) into the wave equation (\ref{Schrodinger}) leads to a four-term recurrence relation 
\beeq
\alpha_1 a_1 +\beta_1 a_0=0
\label{eq-rec1}
\eneq
\beeq
\alpha_2 a_2 +\beta_2 a_1+ \gamma_2 a_0=0
\label{}
\eneq
\beeq
\alpha_n a_n +\beta_n a_{n-1}+ \gamma_n a_{n-2}+ \delta_n a_{n-3}  =0
\label{eq-sevenrecurrence}
\eneq
where $n=3,4,5,\dots$ and $a_0$ is a constant which we can take to be 1.
The coefficients in the above recurrence relation, which are functions of $\omega$, are given in \cite{Leaver-RN} for both gravitational and electromagnetic perturbations.
The QNMs are the values of $\omega$ that satisfy the above four-term recurrence relation for all $n$.
The recurrence relation (\ref{eq-sevenrecurrence}) can be reduced to a three-term recurrence relation 
\beeq
\alpha_n' a_n +\beta_n' a_{n-1}+ \gamma_n' a_{n-2} =0
\eneq
using Gaussian elimination.  For details, see \cite{Leaver-RN}.

Following Leaver in \cite{Leaver}, this three-term recurrence relation gives a continued fraction equation
\begin{eqnarray}
	\beta_1^{'} &=& \frac{\alpha_1^{'} \gamma_{2}^{'}}{\beta_{2}^{'}-\frac{\alpha_{2}^{'}\gamma_{3}^{'}}{\beta_{3}^{'} - \cdots}} \nonumber \\
	&=& \frac{\alpha_1^{'}\gamma_{2}^{'}}{\beta_{2}^{'}-}\frac{\alpha_{2}^{'}\gamma_{3}^{'}}{\beta_{3}^{'} -}\frac{\alpha_{3}^{'}\gamma_{4}^{'}}{\beta_{4}^{'} -}\cdots 
	\label{eqLeaver}
\end{eqnarray}
which can be solved for $\omega$.  For higher overtone QNMs, we use inversions of  (\ref{eqLeaver}) as described in \cite{Leaver}. To evaluate the continued fraction numerically, we  truncate it at some point and approximate the tail-end using Nollert's technique \cite{Nollert}.  

In the extremal case, where $q=M$ ($r_+=r_-$), we have a different topology that requires a new approach suggested by \cite{Onozawa1}.  We review the extremal case in the next section.

\sxn{The Extremal Case}
\vskip 0.3cm

In the extremal case, where $M=q$, the tortoise coordinate becomes
\beeq
r_*=r- \frac{M^2}{r-M}+2M\ln(r-M)~.
\label{}
\eneq
This leads to irregular singularities of the radial wave equation at the event horizon, $r=M$, and at infinity. 
Leaver's method would be to expand the solution around the horizon using $u = (r-M)/r$ in  (\ref{eq-series}).  However, since $r=M$ is an irregular singularity of the differential equation, this will not work since the series will not converge.  Onozawa et al.\ \cite{Onozawa1} gets around this issue by using $u=(r-2M)/r$ instead.  With this choice, the horizon is at $u=-1$ and infinity is at $u=1$.  
 After substituting $\sum_{n} a_n u^n$ into the wave equation, we obtain a five-term recurrence relation for the $a_n$ with coefficients $\alpha_n$,  $\beta_n$,  $\gamma_n$, $\delta_n$, and  $\epsilon_n$. 
Now, convergence of the wavefunction (Eq.\ (17) in \cite{Onozawa1}) at the event horizon and at infinity is equivalent to convergence of  $\sum (-1)^n a_n$  and $\sum a_n$ respectively.  This is the same as requiring $\sum_{n \, \text{odd}} a_n$ and  $\sum_{n \, \text{even}} a_n$ to converge.  
Onozawa shows that the $a_{n\, \text{odd}}$ themselves satisfy a five-term recurrence relation with coefficients 
  $\hat{\alpha}_n$,  $\hat{\beta}_n$,  $\hat{\gamma}_n$, $\hat{\delta}_n$, and  $\hat{\epsilon}_n$, as do the
 $a_{n\, \text{even}}$ with coefficients 
 $\bar{\alpha}_n$,  $\bar{\beta}_n$,  $\bar{\gamma}_n$, $\bar{\delta}_n$, and  $\bar{\epsilon}_n$.  
  These coefficients are related by the following equation\cite{Onozawa1}:
\beeq
\frac{a_3}{a_1}=\frac{ -\dfrac{\alpha_5}{\bar{\alpha}_2''} \bar{\gamma}_2'' + \epsilon_5+ \left( -\dfrac{\alpha_5}{\bar{\alpha}_2''} \bar{\beta}_2'' + \gamma_5\right)\dfrac{a_4}{a_2}  }{\alpha_3 \dfrac{a_4}{a_2}+\gamma_3-\epsilon_3 \dfrac{\alpha_1}{\gamma_1}} ~,
\label{EqQNMEQ}
\eneq
where $a_4/a_2$ and $a_3 / a_1$ can be written as continued fractions of the form
\beeq
 \frac{a_4}{a_2} =   \frac{- \bar{\gamma}_{2}''}{\bar{\beta}_{2}''-\dfrac{\bar{\alpha}_{2}''\bar{\gamma}_{3}''}{\bar{\beta}_{3}'' -\dfrac{\bar{\alpha}_{3}''\bar{\gamma}_{4}''}{\bar{\beta}_{4}''- \cdots}}} 
\eneq
\beeq
\frac{a_3}{a_1} = \frac{- \hat{\gamma}_{1}''}{\hat{\beta}_{1}''-\dfrac{\hat{\alpha}_{1}''\hat{\gamma}_{2}''}{\hat{\beta}_{2}'' -\dfrac{\hat{\alpha}_{2}''\hat{\gamma}_{3}''}{\hat{\beta}_{3}''- \cdots}}} 
\eneq
Here, $''$ indicates the coefficients obtained when the even and odd recurrence relations are reduced to three terms, see \cite{Onozawa1}.\footnote{Note that the exact form of these coefficients in \cite{Onozawa1} require $M=1$.}


The solutions $\omega$ to Eq.\ (\ref{EqQNMEQ}) are the QNMs.  However, using this equation, we are only able to find a few of the solutions.  Typically, when working with continued fraction equations, it is possible to form inversions that make certain roots more numerically stable.  Therefore, the inversion allows one to find other roots. For most black holes, the QNM equation involves one continued fraction, whereas here there are two, $a_4/a_2$ and $a_3/a_1$.   To be able to calculate the QNMs, we consider two different inversion methods.  The first method involves evaluating one continued fraction (either $a_4/a_2$ or $a_3/a_1$) to a fixed depth and inverting the other one.  In the second method, we invert both simultaneously. We briefly explain these two methods below. 

In the first method, we invert the following equation:
\beeq
\frac{- \hat{\gamma}_{1}''}{\hat{\beta}_{1}''-\dfrac{\hat{\alpha}_{1}''\hat{\gamma}_{2}''}{\hat{\beta}_{2}'' -\dfrac{\hat{\alpha}_{2}''\hat{\gamma}_{3}''}{\hat{\beta}_{3}''- \cdots}}} =\mathcal{F}(\omega)
\label{Eq1stInv}
\eneq
where $\mathcal{F}(\omega)$ is the right-hand-side of (\ref{EqQNMEQ}).
We call (\ref{Eq1stInv}) the 1st inversion.
The $n$-th inversion, for $n=2, 3, \dots$, of  (\ref{Eq1stInv}) is defined to be:
\beeq
\hat{\beta}''_n - \frac{\hat{\alpha}''_{n-1}\hat{\gamma}''_n}{\hat{\beta}''_{n-1} -}\, \frac{\hat{\alpha}''_{n-2} \hat{\gamma}''_{n-1}}{\hat{\beta}''_{n-2} -}\cdots \frac{\hat{\alpha}''_1\hat{\gamma}''_2}{\hat{\beta}''_1-}\, \frac{-\hat{\gamma}''_1}{\textcolor{white}{\hat{\textcolor{black}{\mathcal{F}(\omega)}}} }
=
\frac{\hat{\alpha}''_{n} \hat{\gamma}''_{n+1}}{\hat{\beta}''_{n+1} -}  \, \frac{\hat{\alpha}''_{n+1} \hat{\gamma}''_{n+2}}{\hat{\beta}''_{n+2} -} \, \frac{\hat{\alpha}''_{n+2}\hat{\gamma}''_{n+3}}{\hat{\beta}''_{n+3}-}\cdots~.
\label{eqInv}
\eneq
For each inversion we find a number of QNMs.  A rough rule of thumb is that the $n^\text{th}$ QNM, $\omega_n$, is the most stable solution of the $n^\text{th}$ inversion.  Thus, we are able to find a large number of roots by taking more and  more inversions.
Alternatively, we can first solve (\ref{EqQNMEQ}) for $a_4/a_2$ and repeat the above procedure.

Whichever continued fraction is inverted, we find the same roots up to roughly $\omega_{200}$.  Interestingly, after $n\approx 200$, the inversions of Eq.\ (\ref{Eq1stInv}) only give every other root.  Similarly, the equation we get by first solving for $a_4/a_2$ and then inverting also starts to give us every other root.  Conveniently, the two inversion methods are complimentary in finding the roots that are missing in the other one.  Thus, we are able to find roots up to roughly $\omega_{1000}$.  

In the second method, we simultaneously invert both continued fractions, $a_3/a_1$ and $a_4/a_2$.
 To abbreviate the procedure, we rewrite Eq.\ (\ref{EqQNMEQ}) as
\beeq
{\cal{O}}_1=\frac{ -\dfrac{\alpha_5}{\bar{\alpha}_2''} \bar{\gamma}_2'' + \epsilon_5+ \left( -\dfrac{\alpha_5}{\bar{\alpha}_2''} \bar{\beta}_2'' + \gamma_5\right){\cal{E}}_1  }{\alpha_3 {\cal{E}}_1+\gamma_3-\epsilon_3 \dfrac{\alpha_1}{\gamma_1}} ~,
\label{}
\eneq
We can put this in the form
\beeq
A_1 {\cal{O}}_1 + B_1 {\cal{E}}_1+ C_1 {\cal{E}}_1 {\cal{O}}_1 +  D_1 = 0~,
\label{Eq:inversionEq}
\eneq
where
\beeq
{\cal{E}}_1 = \frac{a_4}{a_2}= \frac{- \bar{\gamma}_{2}''}{\bar{\beta}_{2}''-\dfrac{\bar{\alpha}_{2}''\bar{\gamma}_{3}''}{\bar{\beta}_{3}'' - \cdots}} 
\eneq
\beeq
{\cal{O}}_1 = \frac{a_3}{a_1}= \frac{- \hat{\gamma}_{1}''}{\hat{\beta}_{1}''-\dfrac{\hat{\alpha}_{1}''\hat{\gamma}_{2}''}{\hat{\beta}_{2}'' - \cdots}} 
\eneq
\beeq
A_1=  \gamma_3 -\frac{\alpha_1 \epsilon_5}{\gamma_1}~,
\label{}
\eneq
\beeq
B_1=  \frac{\alpha_5}{\bar{\alpha}_2''}\bar{\beta}_2'' - \gamma_5~,
\label{}
\eneq
\beeq
C_1=  \alpha_3~,
\label{}
\eneq
\beeq
D_1=  \frac{\alpha_5}{\bar{\alpha}_2''}\bar{\gamma}_2'' - \epsilon_5~,
\label{}
\eneq
We create ``inversions'' of (\ref{Eq:inversionEq}) by multiplying by the denominator of the continued fractions of ${\cal{E}}_1$ and ${\cal{O}}_1$.  Repeating this process, the $n$th inversion will have the form
\beeq
A_n {\cal{O}}_n + B_n {\cal{E}}_n+ C_n {\cal{E}}_n {\cal{O}}_n +  D_n = 0~,
\label{EqSimulInvert}
\eneq
where

\beeq
{\cal{E}}_n =  \bar{\beta}_{n+1}''- \frac{\bar{\alpha}_{n+1}'' \bar{\gamma}_{n+2}''}{\bar{\beta}_{n+2}''-\dfrac{\bar{\alpha}_{n+2}''\bar{\gamma}_{n+3}''}{\bar{\beta}_{n+3}'' - \cdots}} 
\eneq

\beeq
{\cal{O}}_n = \hat{\beta}_{n}''- \frac{\hat{\alpha}_{n}'' \hat{\gamma}_{n+1}''}{\hat{\beta}_{n+1}''-\dfrac{\hat{\alpha}_{n+1}''\hat{\gamma}_{n+2}''}{\hat{\beta}_{n+2}'' - \cdots}} 
\eneq

\beeq
A_2=  -\bar{\gamma}_2''B_1~,
\label{}
\eneq

\beeq
B_2=  -\hat{\gamma}_1''A_1~,
\label{}
\eneq
\beeq
C_2=  D_1~,
\label{}
\eneq
\beeq
D_2=   \hat{\gamma}_1''\bar{\gamma}_2''C_1~,
\label{}
\eneq

\beeq
A_n=  -\bar{\alpha}_{n-1}''\bar{\gamma}_n''(B_{n-1}+\hat{\beta}_{n-2}'' C_{n-1})~,~~~ n>2
\label{}
\eneq

\beeq
B_n=  \hat{\alpha}_{n-2}''\hat{\gamma}_{n-1}''(A_{n-1}+\bar{\beta}_{n-1}'' C_{n-1})~,~~~ n>2
\label{}
\eneq
\beeq
C_n= D_{n-1}+ \bar{\beta}_{n-1}''B_{n-1} +\hat{\beta}_{n-2}'' (A_{n-1}+\bar{\beta}_{n-1}'' C_{n-1})~,~~~ n>2
\label{}
\eneq
\beeq
D_n=  \hat{\alpha}_{n-2}'' \bar{\alpha}_{n-1}''\hat{\gamma}_{n-1}''\bar{\gamma}_{n}''C_{n-1}~,~~~ n>2
\label{}
\eneq
For each $n$ we can solve Eq.\ (\ref{EqSimulInvert}) for $\omega$.  This procedure gives us the same roots as the previous method. However, we find that it is not as efficient in finding the roots.  Therefore, the results presented in this paper only use the first method.

\begin{figure}[th!]
	\begin{center}
		\includegraphics[height=5.cm]{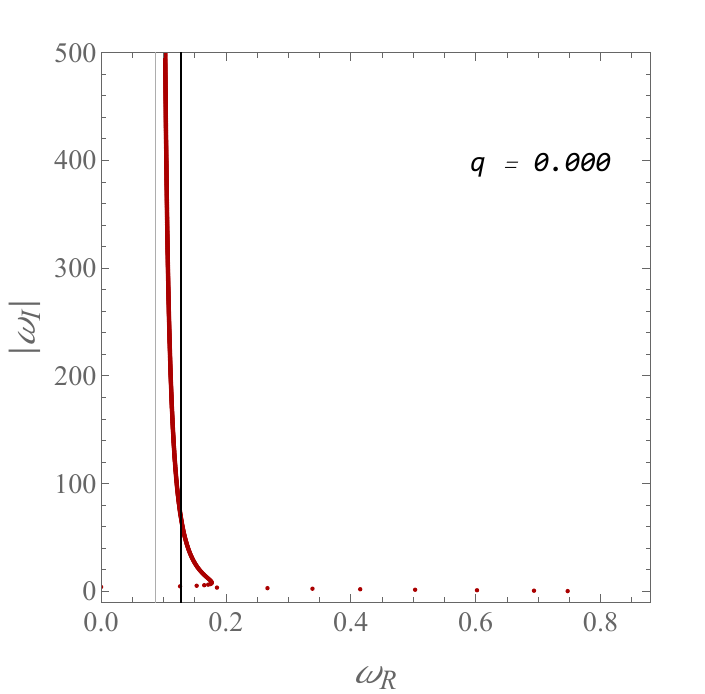}
		\includegraphics[height=5.cm]{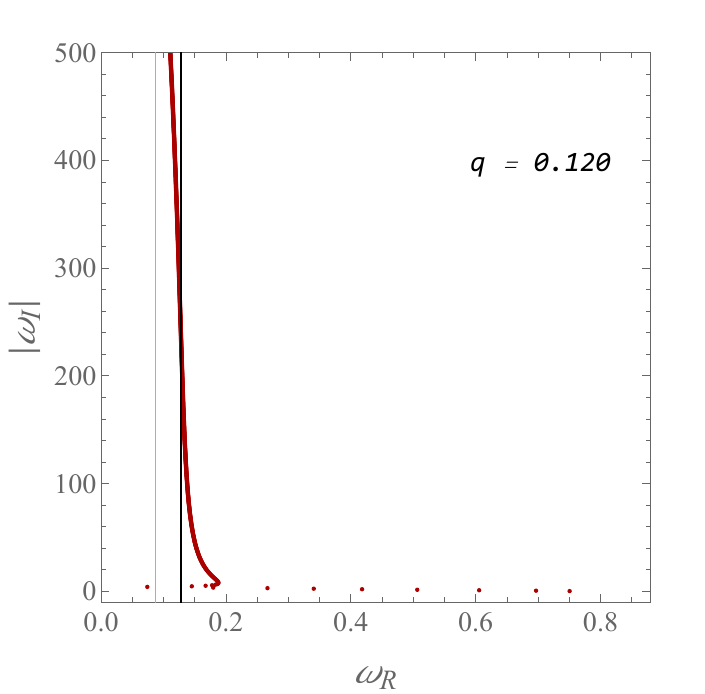}
		\includegraphics[height=5.cm]{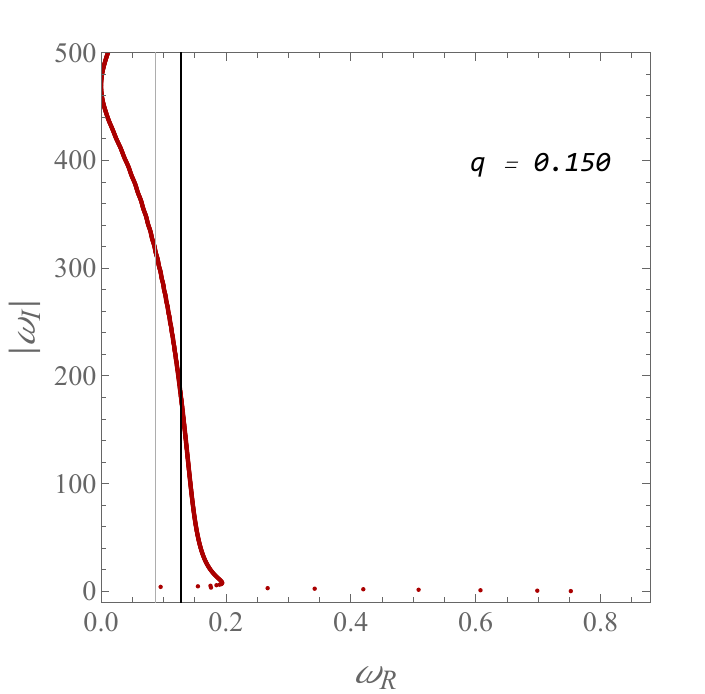}
		\includegraphics[height=5.cm]{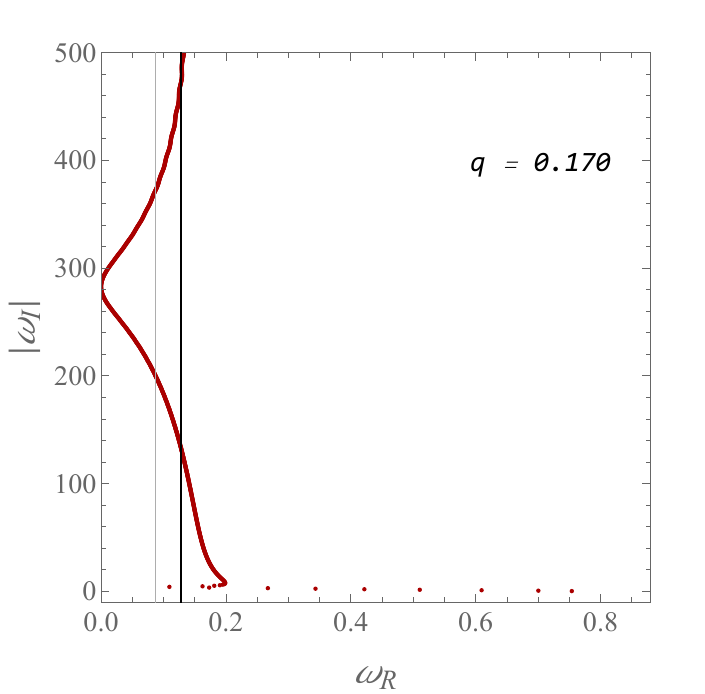}
		\includegraphics[height=5.cm]{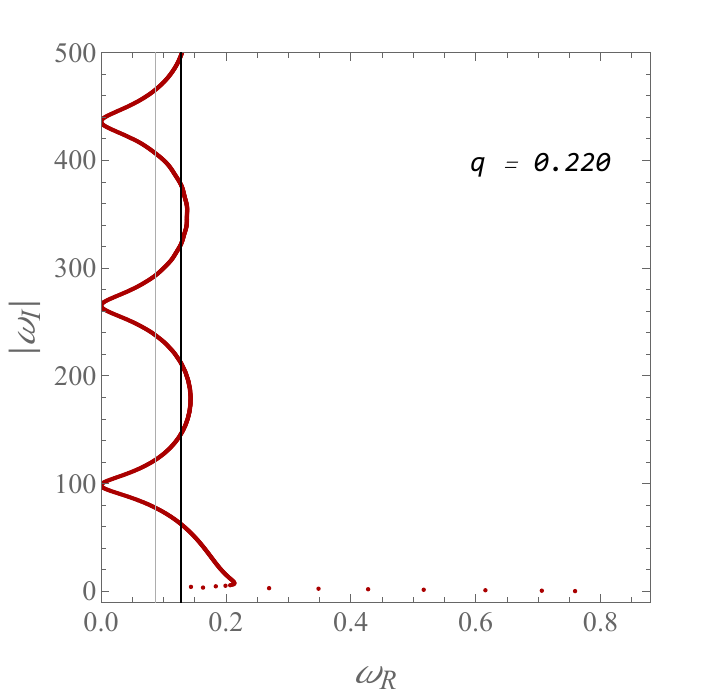}
		\includegraphics[height=5.cm]{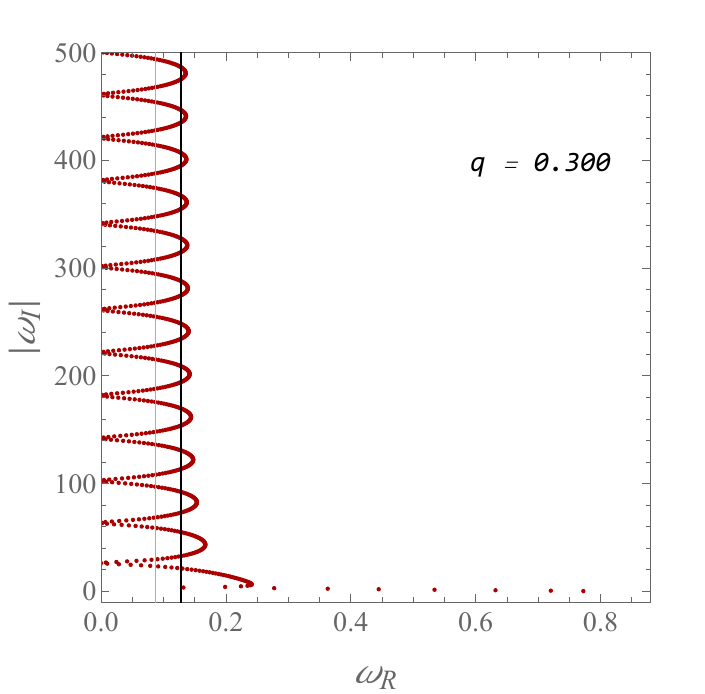}
		\includegraphics[height=5.cm]{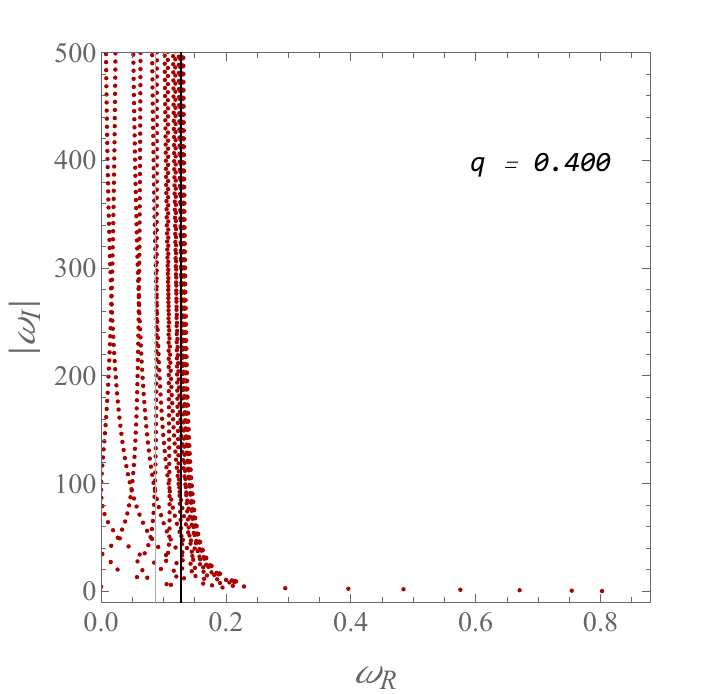}
		\includegraphics[height=5.cm]{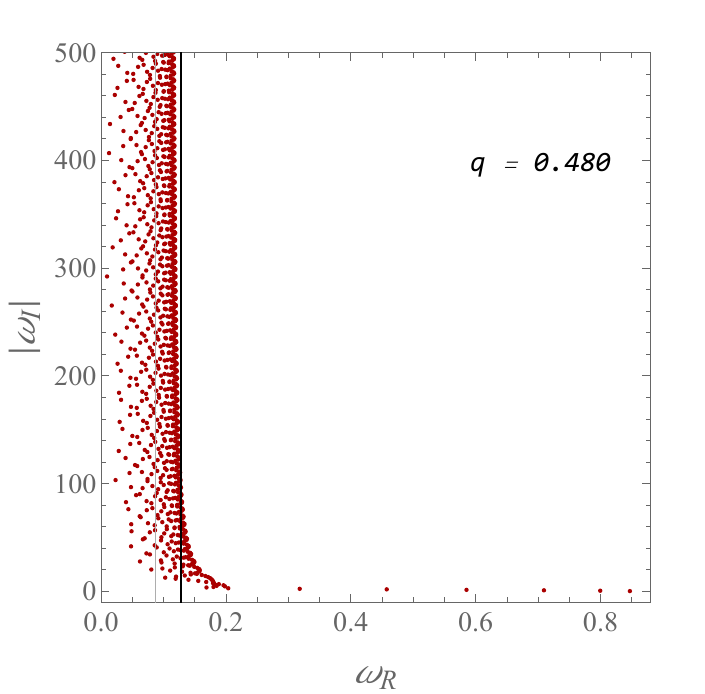}
		\includegraphics[height=5.cm]{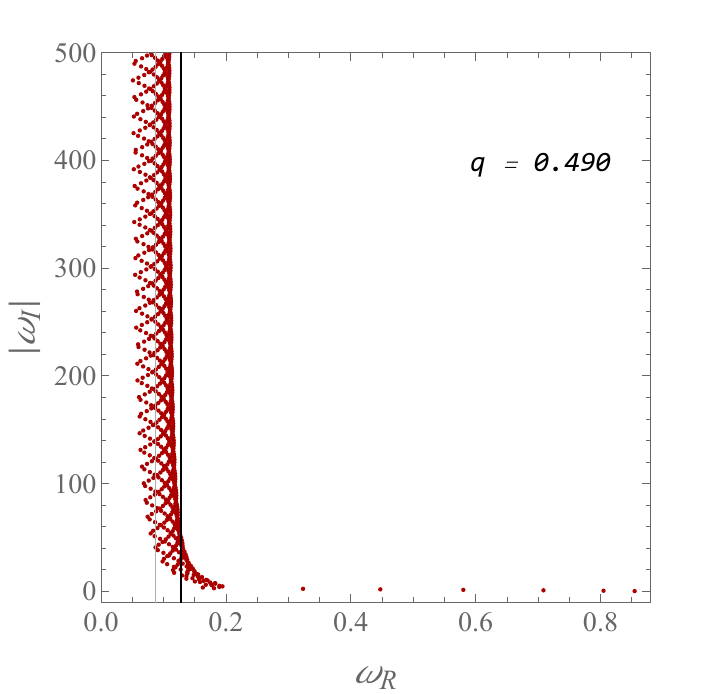}
		\includegraphics[height=5.cm]{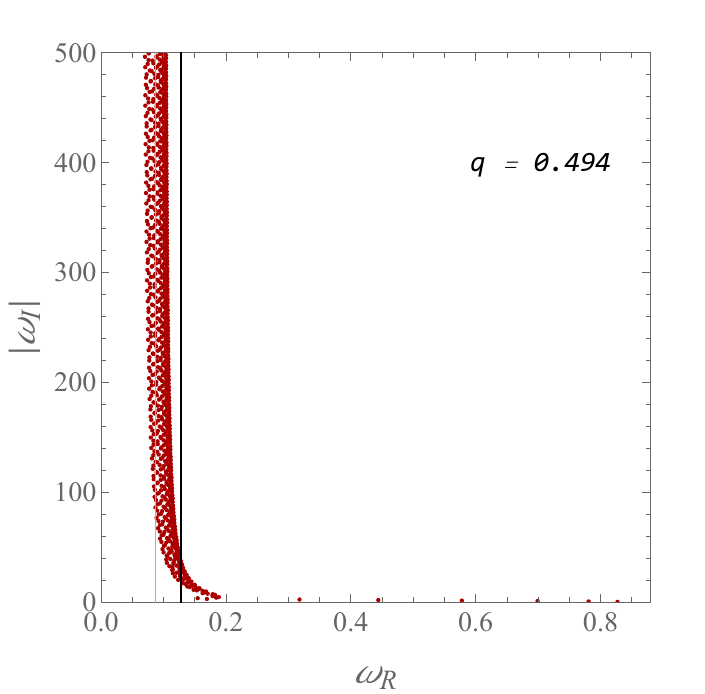}
		\includegraphics[height=5.cm]{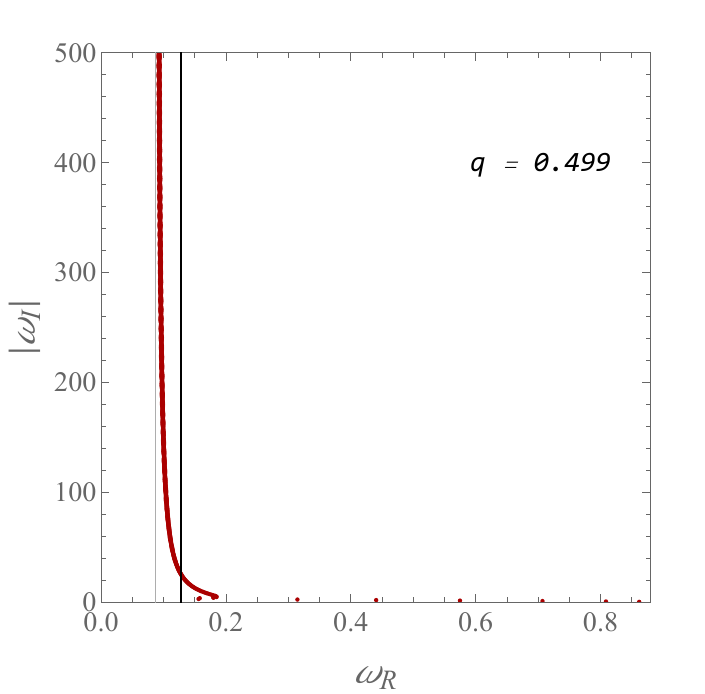}
		\includegraphics[height=5.cm]{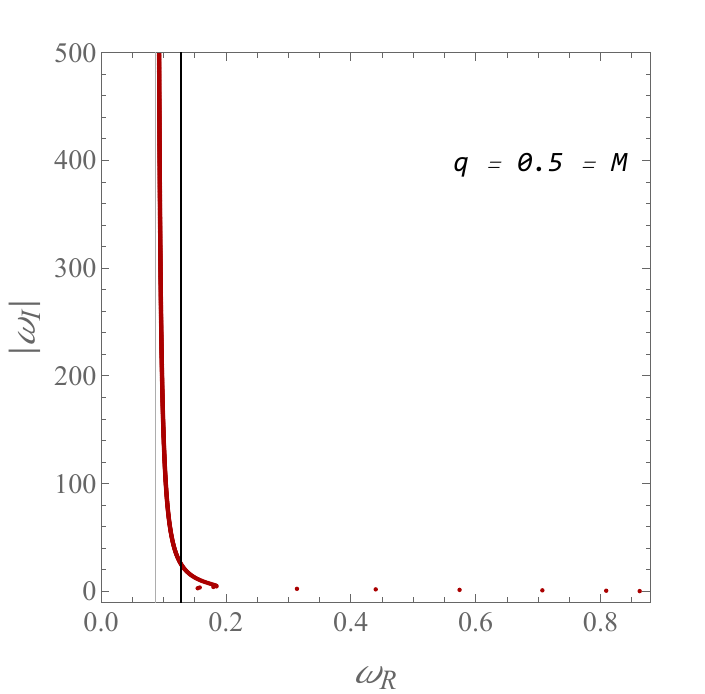}
	\end{center}
	\vspace{-0.7cm}
	\caption{\footnotesize The QNM spectra for different values of charge with $M=0.5$ and $\ell=2$.  The light vertical line is at  $\omega_R=\ln(3)/(8\pi M)$ and the dark vertical line is at  $\omega_R=\ln(5)/(8\pi M)$. }
	\label{fig: QNMspectra}
\end{figure}


\sxn{Numerical Results}
\vskip 0.3cm

Roots are calculated by Leaver's method\cite{Leaver} (with Nollert's improvement\cite{Nollert}) and are verified by checking that the roots remain stable for different depths  in the continued fraction and persist for at least two inversions.


In Fig.\ \ref{fig: QNMspectra}, we show the QNM spectra for various values of charge $q$.  For all charges there is an initial ``bounce" off the imaginary axis at around the ninth mode. Then, for lower values of the charge, there are subsequent bounces starting at some damping rate that decreases with increasing charge.\footnote{For most cases, the bounces do not actually hit the imaginary axis.} The rough overall picture is that the bounces become more frequent as the charge increases and start lifting away from the imaginary axis.  In Fig.\ \ref{fig: QNMspectra}, this lifting becomes visible at $q=0.48$.  This lifting continues until there is no bouncing at the extremal case. 
As the charge approaches the extremal limit, the peaks in between the bounces approach $\ln(3)/(8\pi M)$ as damping increases.  
When the charge approaches zero, the peaks approach $\ln(5)/(8\pi M)$.  

To study the asymptotic behavior of the extremal QNM spectrum, we fit the points $(x,y)=(\omega_I,\omega_R)$ with functions of the form
\beeq
y(x)=\frac{a_0+a_1 x+\dots+a_n x^n}{b_0+b_1 x+\dots+b_n x^n}~,
\label{}
\eneq
We then take the limit as $x\rightarrow -\infty$ [i.e. $y(x\rightarrow -\infty)=a_n/b_n$] to find the asymptotic value of $\omega_R$.   For the extremal case, we found that the best fit is a seventh order rational function and the limit is within $0.0009$ of $\ln(3)/(8\pi M)$. From Fig. \ref{fig: QNMspectra}, one can see that the peaks of the bounces make a transition from   $\ln(5)/(8\pi M)$ to $\ln(3)/(8\pi M)$ as $q$ approaches $M$.
We looked at the asymptotic behavior of the peak of the bounces for $q=0.35$ and $0.45$.  For $q=0.45$, we found that the best fit is a second order rational function and the limit is $0.008$ less than $\ln(5)/(8\pi M)$.  
For $q=0.35$, we found that the best fit is a fourth order rational function and the limit is $0.0003$ more than $\ln(5)/(8\pi M)$.  Therefore, it appears that the peaks of the bounces make a transition from $\ln(5)/(8\pi M)$ to $\ln(3)/(8\pi M)$ in the charge range of $0.35 \lesssim q < 0.5$.  This is consistent with the results in \cite{Berti1}.

Our results show that, as the charge gets smaller, the second bounce occurs at approximately $2 M^3/q^4$, which is consistent with Eq.\ (\ref{Eq: range-small-q}).  However, we cannot determine if for very small charges there are no subsequent bounces in transitioning from $\ln(3)/(8\pi M)$ to $\ln(5)/(8\pi M)$ as  predicted by the semi-analytic calculations of \cite{Daghigh-RN}.

\begin{figure}[th!]
	\begin{center}
		\includegraphics[height=11.2cm]{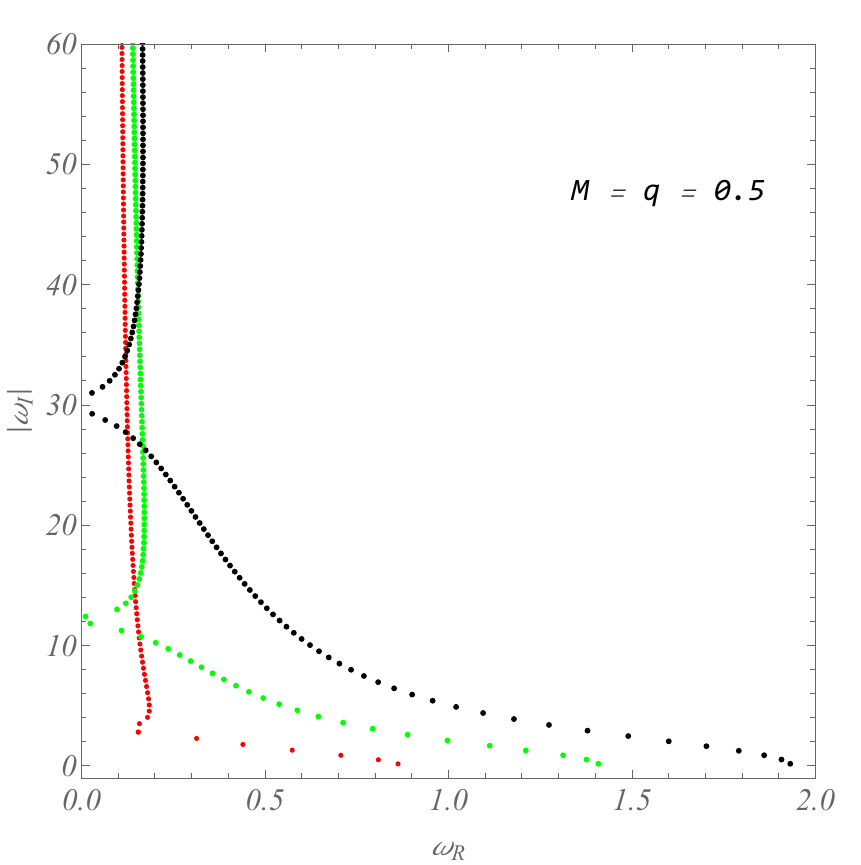}
	\end{center}
	\vspace{-0.7cm}
	\caption{\footnotesize The QNM low damped spectra in the extremal case for, from left to right, $\ell=2$, $3$, $4$. }
	\label{fig: QNMspectra-Extremal-close}
\end{figure}

\begin{figure}[th!]
	\begin{center}
		\includegraphics[height=5.2cm]{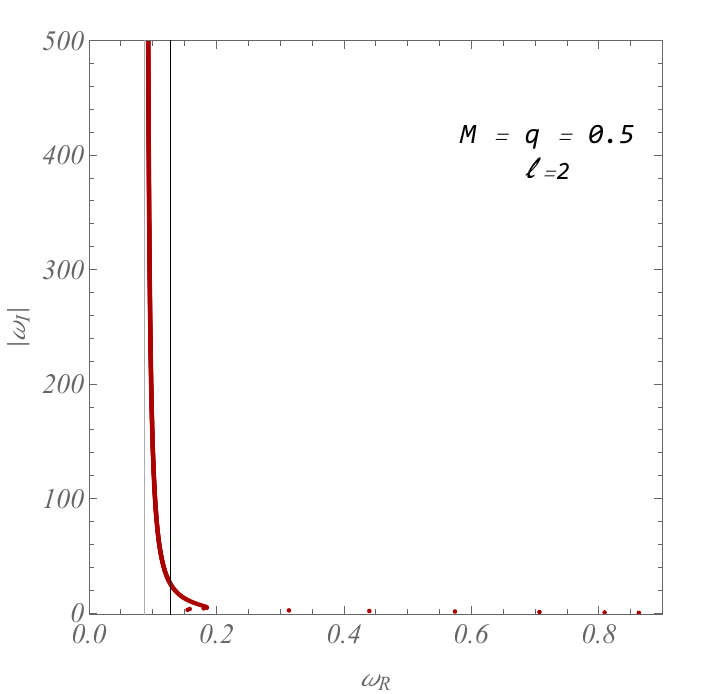}
		\includegraphics[height=5.2cm]{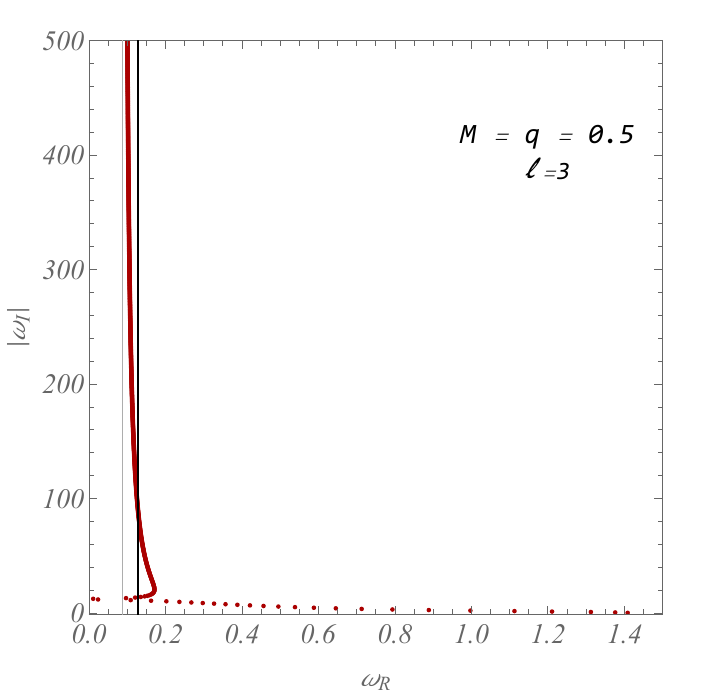}
		\includegraphics[height=5.2cm]{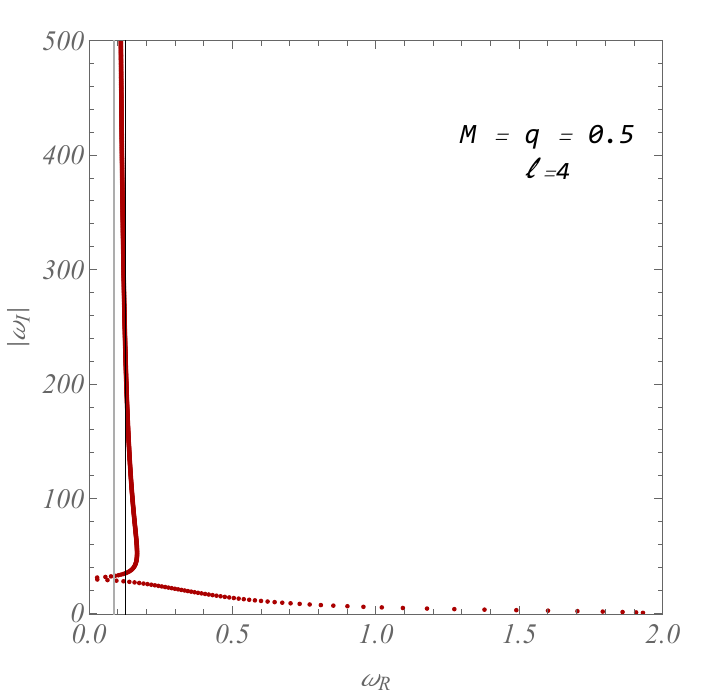}
	\end{center}
	\vspace{-0.7cm}
	\caption{\footnotesize The QNM spectra in the extremal case for, from left to right, $\ell=2$, $3$, $4$.  The light vertical line is at  $\omega_R=\ln(3)/(8\pi M)$ and the dark vertical line is at  $\omega_R=\ln(5)/(8\pi M)$. }
	\label{fig: QNMspectra-Extremal}
\end{figure}

We also looked at the extremal case for $\ell=3$ and $4$.  The results are shown in Fig. \ref{fig: QNMspectra-Extremal-close}, \ref{fig: QNMspectra-Extremal}, and Table I.  Again, to determine the asymptotic value we fit a rational function to the data and found the limit to be within $0.002$ of $\ln(3)/(8\pi M)$ for $\ell=3$ and within $0.004$ for $\ell=4$.

\vspace{0.5cm}
\footnotesize
\begin{tabular}{cccc}
	\multicolumn{4}{c} {Table I.  Gravitational QNMs  for  the extremal case $q=M=0.5$ and different values of $\ell$} \\
	\hline \hline
	$n$ & $\ell = 2$ & $\ell = 3$ & $\ell = 4$ \\
	\hline
1 & $ 0.8626816-0.1669206 i $ & $ 1.408445-0.1717481 i $ & $
1.931525-0.1740026 i $ \\
2 & $ 0.8090471-0.5099687 i $ & $ 1.376931-0.5192489 i $ & $
1.907624-0.5242399 i $ \\
3 & $ 0.7068025-0.8827452 i $ & $ 1.307249-0.8362094 i $ & $
1.860405-0.8812876 i $ \\
4 & $ 0.5742327-1.304909 i $ & $ 1.210897-1.279042 i $ & $
1.791303-1.249989 i $ \\
5 & $ 0.4398891-1.777137 i $ & $ 1.112650-1.673036 i $ & $
1.703111-1.635329 i $ \\
6 & $ 0.3137466-2.280441 i $ & $ 0.9974355-2.117280 i $ & $
1.600492-2.041800 i $ \\
7 & $ 0.1547576-2.806363 i $ & $ 0.8886952-2.590614 i $ & $
1.489995-2.472200 i $ \\
8 & $ -3 i $ (?) & $ 0.7935161-3.084178 i $ & $
1.379015-2.926427 i $ \\
9 & $ 0.1580074-3.505822 i $ & $ 0.7130847-3.589352 i $ & $
1.273919-3.401330 i $ \\
10 & $ 0.1798750-4.033519 i $ & $ 0.6455107-4.100264 i $ & $
1.178613-3.891979 i $ \\
25 & $ 0.1537770-11.63929 i $ & $ 0.0238001-11.83155 i $ & $
0.5586342-11.56845 i $ \\
50 & $ 0.1291992-24.18044 i $ & $ 0.1689169-24.58826 i $ & $
0.2298217-24.21602 i $ \\
100 & $ 0.1135335-49.20611 i $ & $ 0.1459628-49.64689 i $ & $
0.1671020-48.56283 i $ \\
250 & $ 0.1014775-124.2261 i $ & $ 0.1204621-124.6934 i $ & $
0.1446396-123.6476 i $ \\
500 & $ 0.0961954-249.2349 i $ & $ 0.1083683-249.7141 i $ & $
0.1248627-248.6851 i $ \\
750 & $ 0.0941383-374.2385 i $ & $ 0.1033898-374.7226 i $ & $
0.1162330-373.7004 i $ \\
1000 & $ 0.0929643-499.2405 i $ & $ 0.1005974-499.7273 i $ & $
0.1112799-498.7090 i $ \\
	\hline
\end{tabular}
\vspace{0.5cm}
\normalsize

Modes with $\omega_R=0$ are called algebraically special modes and were discovered analytically by Chandrasekhar \cite{Chandra-AS}. They can be found  using the formula given by Berti in \cite{Berti1}\footnote{This corrects a typo in Eq.\ (24) of \cite{Berti1}.  }
\beeq
\omega=\pm i \frac{(\ell-1)\ell (\ell+1)(\ell+2)}{2 [3M+(-1)^s \sqrt{9M^2 +4q^2(\ell-1)(\ell+2)}]}~. 
\label{}
\eneq
For gravitational perturbations, where $s=2$, the above equation gives $\omega_{\ell=2}=-3i $,  $\omega_{\ell=3}=-12i $, and $\omega_{\ell=4}=-30i $ when $M=q=0.5$.    Unfortunately, roots near the imaginary axis are numerically difficult to calculate due to the need for higher depth in the continued fraction Eq\ (\ref{eqInv}). 
For $\omega_{\ell=2}$, Berti found $-3.047876i$, but  we were unable to find this, thus the question mark in Table I.  In Fig. \ref{fig: QNMspectra-Extremal-close}, the gap between neighboring roots for $\omega_{\ell=2}$ seems insufficient to house another root near $-3i$. In the $\ell=3$ case, the closest root we found to the algebraically special frequency was $0.01101539-12.42338 i $. We were unable to find $\omega_{\ell=4}$. However,  in Fig. \ref{fig: QNMspectra-Extremal-close}  there is a big gap between the two neighboring roots  around $\omega_{\ell=4}=-30i$, which suggests there should be another root in this area.    See \cite{Berti1} for discussions regarding the algebraically special frequencies.

\sxn{Conclusions}
\vskip 0.3cm

 Because there has been conflicting results in the literature regarding the asymptotic behavior ($|\omega_I|\rightarrow \infty$) of the QNM spectrum of extremal  Reissner-Nordstr\"om black holes, we numerically calculated over one thousand roots  for multipole indices   $\ell=2$, $3$, and $4$ using a modification of the continued fraction method.   We determined, for all of these indices, the highly damped limit of the real part of the QNMs in the extremal case to be $\ln(3)/(8\pi M)$, consistent with the extremal limit of the non-extremal case.  This agrees with the analytic result found in \cite{DG-extremal}.
 
 We also examined the behavior of the spectrum in the non-extremal case for $\ell=2$.  Here, the spectrum has a ``bouncing" behavior.  The peaks of the bounces asymptotically appear to approach  $\ln(5)/(8\pi M)$ for small values of charge, but then transition to  $\ln(3)/(8\pi M)$ as the charge becomes larger.  In the limit where charge goes to zero, \cite{Daghigh-RN} shows that the spectrum approaches  $\ln(3)/(8\pi M)$ in an intermediate damping region and then bounces once and transitions to  $\ln(5)/(8\pi M)$ in the high damping limit.  While our results are consistent with this behavior, we were not able to compute enough roots to confirm it.

\vspace{1cm}
\leftline{\bf Acknowledgments}
We thank Metropolitan State University for providing a powerful computer that was used extensively for the calculations in this paper.


\def\jnl#1#2#3#4{{#1}{\bf #2} (#4) #3}

\def\Zphys{{\em Z.\ Phys.} }
\def\jssc{{\em J.\ Solid State Chem.} }
\def\jpsJ{{\em J.\ Phys.\ Soc.\ Japan} }
\def\ptps{{\em Prog.\ Theoret.\ Phys.\ Suppl.\ } }
\def\PTP{{\em Prog.\ Theoret.\ Phys.\  }}
\def\LNC{{\em Lett.\ Nuovo.\ Cim.\  }}
\def\LRR{{\em Living \ Rev.\ Relative.} }
\def\JMP{{\em J. Math.\ Phys.} }
\def\NPB{{\em Nucl.\ Phys.} B}
\def\NP{{\em Nucl.\ Phys.} }
\def\PLB{{\em Phys.\ Lett.} B}
\def\PL{{\em Phys.\ Lett.} }
\def\PRL{\em Phys.\ Rev.\ Lett. }
\def\PRB{{\em Phys.\ Rev.} B}
\def\PRD{{\em Phys.\ Rev.} D}
\def\PRX{{\em Phys.\ Rev.} X~}
\def\PR{{\em Phys.\ Rev.} }
\def\PRe{{\em Phys.\ Rep.} }
\def\AP{{\em Ann.\ Phys.\ (N.Y.)} }
\def\RMP{{\em Rev.\ Mod.\ Phys.} }
\def\ZPC{{\em Z.\ Phys.} C}
\def\SCI{\em Science}
\def\CMP{\em Comm.\ Math.\ Phys. }
\def\MPLA{{\em Mod.\ Phys.\ Lett.} A}
\def\IJMPB{{\em Int.\ J.\ Mod.\ Phys.} B}
\def\cmp{{\em Com.\ Math.\ Phys.}}
\def\JPA{{\em J.\  Phys.} A}
\def\CQG{\em Class.\ Quant.\ Grav.~}
\def\ATMP{\em Adv.\ Theoret.\ Math.\ Phys.~}
\def\PRSA{{\em Proc.\ Roy.\ Soc.\ Lond.} A }
\def\IJTP{\em Int.\ J.\ Theor.\ Phys.~}
\def\GERG{\em Gen.\ Rel.\ Grav.~}
\def\JHEP{\em JHEP~}
\def\ibid{{\em ibid.} }

\vskip 1cm

\leftline{\bf References}

\renewenvironment{thebibliography}[1]
        {\begin{list}{[$\,$\arabic{enumi}$\,$]}  
        {\usecounter{enumi}\setlength{\parsep}{0pt}
         \setlength{\itemsep}{0pt}  \renewcommand{\baselinestretch}{1.2}
         \settowidth
        {\labelwidth}{#1 ~ ~}\sloppy}}{\end{list}}


\end{document}